\documentstyle[sprocl,epsfig]{article}

\def\Journal#1#2#3#4{{#1} {\bf #2}, #3 (#4)}

\def\NPB{{\em Nucl. Phys.} B}
\def\PLB{{\em Phys. Lett.}  B}
\def\PRL{\em Phys. Rev. Lett.}
\def\PRD{{\em Phys. Rev.} D}
\def\ZPC{{\em Z. Phys.} C}

\def\be{\begin{equation}}
\def\ee{\end{equation}}
\def\bea{\begin{eqnarray}}
\def\eea{\end{eqnarray}}

\begin{document}

\title{POLARIZED FRAGMENTATION FUNCTIONS}

\author{Daniel de Florian \footnote{This work was supported in part by the EU Fourth Framework Programme `Training and Mobility
of Researchers', Network `Quantum Chromodynamics and the Deep Structure of Elementary Particles', contract FMRX-CT98-0194 (DG 12 - MIHT)}}

\address{Institute of Theoretical Physics, ETH\\ CH-8093 Z\"urich, Switzerland\\E-mail: dflorian@itp.phys.ethz.ch}

\maketitle\abstracts{
In this talk I present a review on the theoretical status of polarized fragmentation functions and  the prospects 
for conceivable future semi-inclusive deep-inelastic scattering  and proton-proton collision experiments to measure them.}

\section{Introduction}
Measurements of rates for single-inclusive $e^+e^-$ annihilation (SIA) into 
a specific hadron $H$,
\begin{equation}
e^+e^- \rightarrow (\gamma,\,Z) \rightarrow H\;X\;\;\;,
\end{equation}
play a similarly fundamental role as those of the corresponding crossed 
`space-like' deep-inelastic scattering (DIS) 
process $ep\rightarrow e' X$. Their 
interpretation in terms of scale-dependent
fragmentation functions $D_f^{H}(z,Q^2)$, the `time-like' counterparts of the 
parton distribution functions $f_H(x,Q^2)$ of a hadron $H$, hence provides
a further important, complementary test of perturbative QCD.
 Furthermore, in the polarized case, they provide information about how the spin carried by the partons is passed to the hadrons in the fragmentation process and, at variance with DIS, it is  possible to measure them even for the case of hadrons which are not available as  targets.

Among all the hadrons, the production of $\Lambda$ baryons appears to be particularly  
interesting: the self-analyzing properties of its dominant weak decay 
$\Lambda \rightarrow p \pi^-$ and the particularly large asymmetry of the 
angular distribution of the decay proton in the $\Lambda$ rest-frame 
allow for an experimental reconstruction of the $\Lambda$ spin. 

In the following we will address the issue of fragmentation into polarized
$\Lambda$'s within a detailed QCD analysis. We will provide some realistic sets of unpolarized and polarized 
fragmentation functions for $\Lambda$ baryons. Since there are hardly 
any data sensitive to the polarized fragmentation 
functions $\Delta D_f^{\Lambda}$,
one has to rely on reasonable assumptions.

Our various proposed sets for 
$\Delta D_f^{\Lambda}$, which are all compatible with the LEP 
data, are 
particularly suited for estimating the physics potential of different processes to 
determine the polarized fragmentation functions 
more precisely. We hence present 
detailed predictions for Semi-Inclusive DIS (SIDIS) and $pp$ collisions, for both longitudinally and transversely polarized beams and $\Lambda$'s.
For further details we refer the reader to the original refs. \cite{lamdis,lampp,lamtrans}.

\section{Unpolarized $\Lambda$ Fragmentation Functions}

The cross section for the inclusive production of a hadron with 
energy $E_H$ in SIA at a c.m.s. energy $\sqrt{s}$, integrated over the 
production angle, can be written in the following 
way \cite{altarelli}:
\begin{equation}
\label{eq:cross1}
\frac{1}{\sigma_{tot}} \frac{d\sigma^H}{dx_E} = \frac{1}{\sum_q e_q^2}
\left[ 2\, F_1^{H}(x_E,Q^2) +  F_L^{H}(x_E,Q^2) \right] \; ,
\end{equation}
where $x_E=2 p_H\cdot k/Q^2 = 2 E_H/\sqrt{s}$ ($k$ being the 
momentum of the intermediate boson, $k^2=Q^2=s$) and
$\sigma_{tot}$ is the the total cross section for 
$e^+e^- \rightarrow hadrons$. The sum in (\ref{eq:cross1}),
runs over the $n_f$ active quark flavors $q$, and 
the $e_q$ are the corresponding appropriate electroweak charges.

The unpolarized `time-like' structure function $F_1^{H}$ is in LO related 
to the fragmentation functions $D_f^H(z,Q^2)$ by 
\begin{eqnarray}
2 \, F_1^{H}(x_E,Q^2) &=& \sum_q e_q^2 \left[ D_q^H (x_E,Q^2) + 
D_{\bar q}^H (x_E,Q^2) \right]\;\;\; ,  
\end{eqnarray}
the $D_f^H (z,Q^2)$ obeying LO Altarelli-Parisi-type $Q^2$-evolution 
equations. The corresponding NLO  expressions for 
$F_1^H$ and $F_L^H$ which include the relevant NLO ($\overline{\mathrm{MS}}$) coefficient functions 
are too lengthy to be given here but can be found in \cite{altarelli}.
In the last few years several experiments \cite{data} have reported 
measurements of the unpolarized cross section for the production of 
$\Lambda$ baryons, which allows an extraction of the unpolarized $\Lambda$ 
fragmentation functions required for constructing the polarization 
asymmetries and as reference distributions in the positivity constraint 
for the polarized ones. We emphasize at this point that the wide range 
of c.m.s.\ energies covered by the data  
($14 \leq \sqrt{s} \leq 91.2$ GeV) makes a detailed QCD analysis that 
includes the $Q^2$-evolution of the fragmentation functions mandatory.

Unless stated otherwise, we will refer to both $\Lambda^0$ and 
$\bar{\Lambda}^0$ as simply `$\Lambda$'. 
As a result, the obtained fragmentation functions always correspond to the sum
\begin{equation}
\label{eq:lamlambar}
D_f^\Lambda(x_E,Q^2) \equiv D_f^{\Lambda^0}(x_E,Q^2)+ 
D_f^{\bar{\Lambda}^0}(x_E,Q^2)\;\;\;.
\end{equation}
 Since no precise 
SIDIS data are available yet, it is not possible to obtain individual 
distributions for all the light flavors separately, and hence some sensible 
assumptions concerning them have to be made. Employing naive quark model 
$SU_f(3)$ arguments and neglecting any mass differences between 
$u,d,s$, we {\it assume} 
that all the light flavors fragment equally into $\Lambda$, i.e.
\begin{eqnarray}
\label{eq:ansatz}
D^{\Lambda}_u=D^{\Lambda}_d=D^{\Lambda}_s=D^{\Lambda}_{\bar{u}}=
D^{\Lambda}_{\bar{d}}=D^{\Lambda}_{\bar{s}} 
\equiv D^{\Lambda}_q\;\;.
\end{eqnarray} 

At the rather low initial scale $\mu$ \cite{grv} ($\mu^2_{LO}=0.23\,{\mathrm{GeV}}^2$, 
$\mu^2_{NLO}=0.34\,{\mathrm{GeV}}^2$) we choose the following simple ansatz:
\begin{equation}
\label{eq:input}
z\, D^{\Lambda}_f (z,\mu^2) = N_f\, z^{\alpha_f} (1-z)^{\beta_f} \; ,
\end{equation}
where $f=u,d,s$, and $g$. For details on the treatment of the 
heavy flavor contributions,  we refer to \cite{lamdis}.
Utilizing eq.\ (\ref{eq:ansatz}) a total of 10 free parameters remains 
to be fixed from a fit to the available 103 data 
points \cite{data} (we include only data with $x_E>0.1$
in our fit to avoid possibly large non-perturbative contributions
induced by finite-mass corrections). The total 
$\chi^2$ values are 103.55 and 104.29 in NLO and LO, respectively, and 
the optimal parameters in (\ref{eq:input}) can be 
found in~\cite{lamdis}. 

A comparison of our LO and NLO results with the data is presented in fig.\ 1, 
where all the existing data \cite{data}
have been converted to the `format' of 
eq.(\ref{eq:cross1}). One should note that the LO and the NLO results are 
almost indistinguishable, demonstrating the perturbative stability of
the process considered. Furthermore, there is an excellent agreement between 
the predictions of our fits and the data even in the small-$x_E$ region 
which has not been included in our analysis.
%
%
\begin{figure}[th]
\begin{center}
\vspace*{-1cm}
\epsfig{file=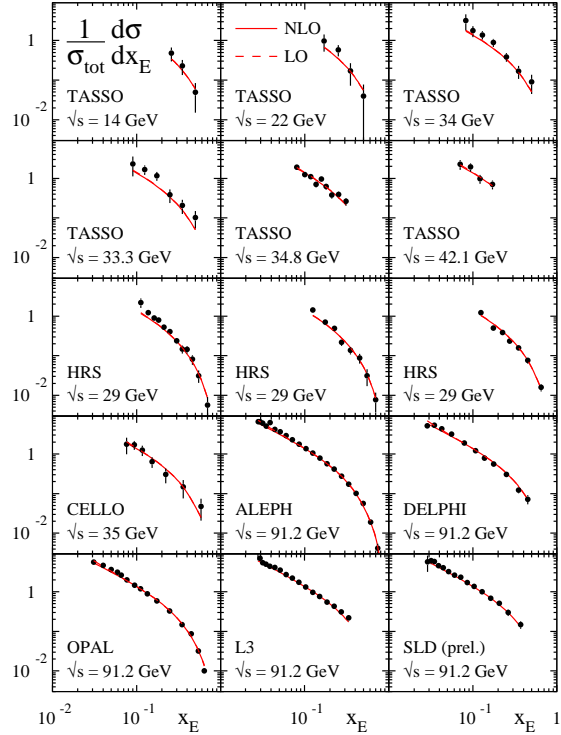,width=8.5cm}
\vspace*{-0.5cm}
\caption{Comparison of LO and NLO results with all available data on 
 unpolarized $\Lambda$ production in $e^+ e^-$ annihilation. 
 Note that only data points with $x_E\ge 0.1$ have been included in our fit.}
\vspace*{-0.3cm}
\end{center}
\end{figure}

\section{Polarized Fragmentation Functions}
%
\noindent 
Having obtained a reliable set of unpolarized fragmentation 
functions we now turn to the polarized case where unfortunately
only scarce and far less precise data are available. In fact, no data at 
all have been obtained so far using polarized beams. The
only available information comes from {\it un}polarized LEP measurements 
 profiting from the parity-violating 
electroweak $q\bar{q}Z$ coupling. 

For such measurements, done at the mass of the $Z$ boson (`$Z$-resonance'), 
the  asymmetry is given by
\begin{equation}
\label{eq:lepasym}
A^H = \frac{ g_3^H + g_L^H/2}{ F_1^H+F_L^H/2}.
\end{equation}
The polarized {\it non}-singlet structure function $g_3^H$ is given at LO by 
\begin{equation}
g_3^H(x_E,Q^2) = \sum_q \, g_q \left[
\Delta D_q^H (x_E,Q^2) -\Delta D_{\bar q}^H (x_E,Q^2) \right]\;\;\;,    
\end{equation}
where the $g_q$ are the appropriate effective charges 
(see, e.g., \cite{lamdis}). The complete NLO QCD corrections 
can be found in \cite{lamdis,ravi}. 
Note that only the {\it valence} part 
of the polarized fragmentation functions can be obtained from the available 
LEP data~\cite{data}, and that $\Lambda^0$'s and 
$\bar{\Lambda}^0$'s give contributions of opposite signs to the measured 
polarization and thus to $g_3^\Lambda$. Unfortunately, 
it turns out that with the available LEP data~\cite{data}, 
all obtained on the $Z$-resonance, it is not 
even possible to obtain the valence distributions for all the flavors, 
so some assumptions have to be made here. Obviously, even further 
assumptions are needed for the polarized gluon and sea fragmentation 
functions in order to have a complete set of fragmentation functions 
suitable for predictions for other processes.

The heavy flavor contributions to polarized $\Lambda$ 
production are neglected, and $u$ and $d$ fragmentation functions 
are taken to be equal in this analysis. Furthermore, polarized  
unfavored distributions, i.e. $\Delta D_{\bar{u}}^{\Lambda^0} = 
\Delta D_{u}^{\bar{\Lambda}^0}$, etc., and the gluon fragmentation 
function $\Delta D_g^\Lambda$ are assumed to be negligible at the 
initial scale $\mu$. The remaining fragmentation functions are 
parameterized in the following simple way
\begin{equation}
\label{eq:polinput}
\Delta D_s^\Lambda (z,\mu^2) = z^\alpha  D_s^\Lambda (z,\mu^2)\;\;,\;\;
\Delta D_u^\Lambda (z,\mu^2) = \Delta  D_d^\Lambda (z,\mu^2) = 
N_u\, \Delta D_s^\Lambda (z,\mu^2) 
\end{equation}
and are subject to  positivity constraints.
These input distributions are then evolved to higher $Q^2$ via the 
appropriate Altarelli-Parisi equations \cite{apmw}. %

Within this framework we try three different scenarios 
for the polarized fragmentation functions at our low
initial scale $\mu$ to cover a rather wide range of plausible models: 

\noindent
{\bf{Scenario 1}} corresponds to the expectations from the non-relativistic 
naive quark model where only $s$-quarks can contribute to the fragmentation
processes that eventually yield a polarized $\Lambda$, even if the $\Lambda$
is formed via the decay of a heavier hyperon. We hence have $N_u=0$ in 
(\ref{eq:polinput}) for this case.

\noindent
{\bf{Scenario 2}} is based on estimates by Burkardt and 
Jaffe \cite{burkjaf,jaffe2} for the DIS structure function 
$g_1^{\Lambda}$ of the $\Lambda$, predicting
sizeable negative contributions from $u$ and $d$ quarks to $g_1^{\Lambda}$ by 
analogy with the breaking of the Ellis-Jaffe sum rule 
for the proton's $g_1^p$. Assuming that such features also carry over
to the `time-like' case \cite{jaffe2}, we simply impose $N_u=-0.20$ 
(see also \cite{bravar}).

\noindent
{\bf{Scenario 3:}} All the polarized fragmentation functions are assumed 
to be equal here, i.e. $N_u=1$, contrary to the expectation  
of the non-relativistic quark model used in scen.\ 1. This rather `extreme' 
scenario might be realistic if, e.g., there are sizeable contributions to
polarized $\Lambda$ production from decays of heavier hyperons who have
inherited the polarization of originally produced $u$ and $d$ quarks.

Our results for the asymmetry $A^{\Lambda}$ in (\ref{eq:lepasym}) within 
the three different scenarios are compared to the available LEP 
data~\cite{data} in fig.\ 2.
The optimal parameters in (\ref{eq:polinput}) for the three models can
be found in \cite{lamdis}. As can be seen, the best agreement with 
the data is obtained within the (naively) most unlikely scen.\ 3.  
The differences occur mainly in the region of large $x_E$, where 
scen.\ 1 and 2 cannot fully account for the rather large 
observed polarization 
due to the positivity constraints. 
For instance, in the case of scen.\ 1, the 
asymmetry behaves asymptotically roughly like $-\Delta D_s^\Lambda / 3D_s^\Lambda$, 
and even when saturating the positivity constraint  
at around $x_E=0.5$ it is not possible to obtain a polarization as 
large as the one required by the LEP data. Of course, such an 
argument still depends strongly on the assumed $SU(3)_f$ symmetry for 
the {\it{un}}polarized fragmentation functions, which could be broken. The 
situation concerning the $\Lambda$ fragmentation functions can only be 
further clarified by future precise  measurements 
for unpolarized {\em and} polarized beams.
%
\begin{figure}[th]
\begin{center}
\vspace*{-0.6cm}
\epsfig{file=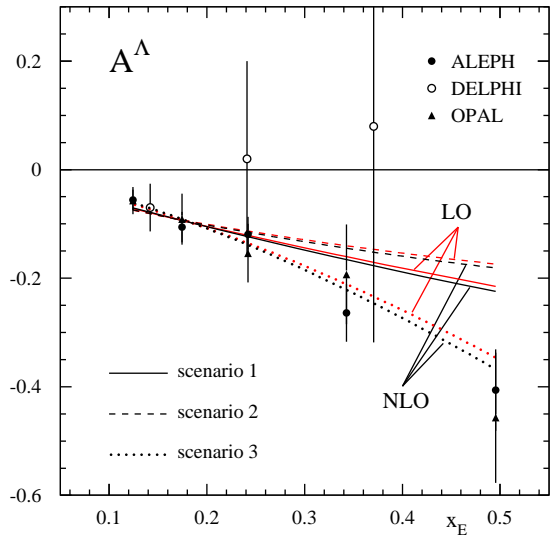,width=8.5cm}
\vspace*{-0.95cm}
\caption{ Comparison of LEP data  and our LO 
and NLO results for the asymmetry $A^{\Lambda}$, using the three different
scenarios.}
\vspace*{-0.1cm}
\end{center}
\end{figure} 

Equipped with various sets of polarized fragmentation 
functions, let us now turn to other processes that might further probe our 
distributions.

%
\section{SIDIS}
\noindent

The SIDIS process $e N \rightarrow e' H X$ \footnote{The case of SIDIS with a neutrino beam has been studied in \cite{js2}.} is  very well suited to 
give information on fragmentation functions. In this case, the cross 
section is proportional to a combination of both the parton distributions 
of the nucleon $N$ and the fragmentation functions for the hadron $H$ (in this work, we will only refer to the current fragmentation region).

In the particular case where both nucleon and hadron are unpolarized, the 
cross section can be written in a way similar to the fully inclusive DIS 
case:
\begin{equation}
\label{eq:sidis}
\frac{d\sigma^H}{dx\, dy\, dz} = 
\frac{4\, \pi\alpha^2 x s}{Q^4} 
\left[ (1+(1-y)^2) F_1^{N/H}(x,z,Q^2) + 
\frac{(1-y)}{x} F_L^{N/H}(x,z,Q^2) \right] 
\end{equation}
with the structure function $F_1^{N/H}$ given at LO by
\begin{equation}
\label{eq:nopol} 
2\, F_{1}^{N/H}(x,z,Q^2) = \sum_{q,\overline{q}} 
e_q^2 \, q (x,Q^2)  D^H_q (z,Q^2)  \; . 
\end{equation}
The corresponding NLO corrections can be found in \cite{graudenz,altarelli}.
Three other possible cross
sections can be defined when the polarization of the lepton, the initial 
nucleon and the hadron are taken into account. 
If both nucleon and hadron are polarized and the lepton is unpolarized, 
the expression is similar to eqs.\ (\ref{eq:sidis}),(\ref{eq:nopol}) 
above with, however, the unpolarized parton distributions {\em and} the 
fragmentation functions to be replaced by their polarized counterparts. 
In the case that the lepton and either 
the nucleon or the hadron are polarized, the expression for the cross 
section is given as in the fully inclusive case by a single structure 
function $g_{1}^{N/H}(x,z,Q^2)$: 
\begin{equation}
\label{eq:sidispol}
\frac{d\Delta\sigma^H}{dx\, dy\, dz} = \frac{8 \pi\alpha^2 x y s}{Q^4} 
\left[ (2-y)\, g_1^{N/H}(x,z,Q^2)  \right]  \; .
\end{equation}
Here the polarized structure function $g_1^{N/H}$ can be written as
\cite{singlepol,doublepol}
\begin{equation}
\label{eq:pol1} 
2\, g_{1}^{N/H}(x,z,Q^2) = \sum_{q,\overline{q}} 
e_q^2 \, (\Delta) q (x,Q^2) (\Delta) D^H_q (z,Q^2)  \; , 
\end{equation}
the position of the $\Delta$ depending on which particle is polarized.
The corresponding NLO corrections  can be found in \cite{singlepol,doublepol,thesis,lamdis}.

The most interesting observable at HERA with respect to the determination
of the polarized $\Lambda$ fragmentation functions is of course the 
asymmetry for the production of {\em polarized} $\Lambda$'s from an 
unpolarized proton, defined by $A^{\Lambda}\equiv g_1^{p/\Lambda}/
F_1^{p/\Lambda}$. Such a measurement 
would be particularly suited 
to improve the information on polarized fragmentation functions. 
In fig.\ 3a), we show our LO and NLO predictions for HERA with polarized 
electrons and {\it{un}}polarized protons using the GRV parton 
distributions \cite{grv}, integrated over the measurable range 
$0.1 \leq z \leq 1$. Good perturbative stability of 
the process is found. As can be seen, the results obtained using the 
three distinct scenarios for polarized fragmentation functions turn 
out to be completely different. 
%
%
\begin{figure}[th]
\begin{center}
\vspace*{-1.6cm}
\hspace*{0.8cm}
\epsfig{file=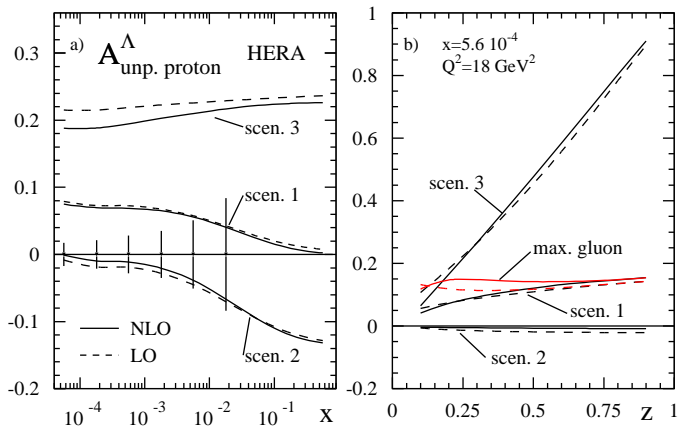,width=8.5cm}
\vspace*{-1.cm}
\caption{\sf LO and NLO predictions for the SIDIS asymmetry for 
unpolarized protons and polarized $\Lambda$'s and leptons 
(see text) for our three
distinct scenarios of polarized fragmentation functions. In {\bf a)} we also
show the expected statistical errors for such a measurement at HERA.}
\vspace*{-0.1cm}
\end{center}
\end{figure}

We have included in fig.~3a) also the expected statistical errors 
for HERA, computed assuming 
an integrated luminosity of $500$ pb$^{-1}$ and a realistic value of
$\epsilon=0.1$ for the efficiency of $\Lambda$ detection. 
Comparing the asymmetries and the error bars in fig.~3a) one concludes that a 
measurement of $A^{\Lambda}$ at small $x$ would allow a discrimination 
between different conceivable scenarios for polarized fragmentation functions.
Fig.~3b) shows our results vs.\ $z$ for fixed $x=5.6 \cdot 10^{-4}$. Again, 
very different asymmetries are found for the three scenarios. 
%
%
\begin{figure}[th]
\begin{center}
\vspace*{-1.2cm}
\epsfig{file=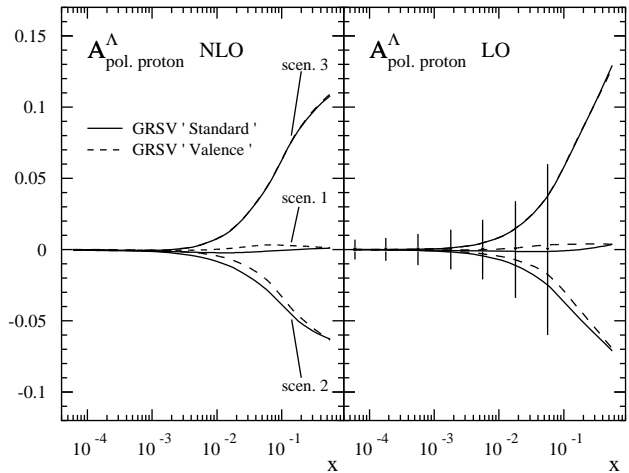,width=8.5cm}
\vspace*{-1.2cm}
\caption{LO and NLO predictions for the SIDIS asymmetry for polarized
protons but unpolarized leptons for two different sets of polarized
parton distributions.}
\vspace*{-0.1cm}
\end{center}
\end{figure}

Interesting results are also obtained for fixed target experiments, where the kinematical range in $x$ and $z$ is complementary to the one of HERA; we refer the reader to \cite{lamdis} for details on that.

The particular case of both target and hadron being polarized was originally 
proposed as a very good way to obtain the $\Delta s$ distribution~\cite{luma}. 
The underlying assumption here was that only the fragmentation function 
$\Delta D_s^{\Lambda}$ is sizeable (as realized, e.g. in our scenario 1), and 
that therefore the only contribution to the polarized cross section has to be 
proportional to $\Delta s \Delta D_s^{\Lambda}$. In order to analyze the 
sensitivity of the corresponding asymmetry to $\Delta s$, we compute it 
using the two different GRSV sets of polarized parton densities of the 
proton \cite{grsv}, which mainly differ in the strange distribution:
the so-called `standard' set assumes an unbroken $SU(3)_f$ symmetric
sea, whereas in the `valence' scenario the sea is maximally broken
and the resulting strange quark density quite small.

The results are shown in fig.\ 4. Unfortunately -- and not unexpectedly --
it turns out that the differences in the asymmetry resulting from our 
different models for polarized $\Lambda$ fragmentation are far larger 
than the ones due to employing different polarized proton strange densities.
In addition, a distinction between different $\Delta s$ would remain 
elusive even if the spin-dependent $\Lambda$ fragmentation functions
were known to good accuracy, as can be seen from the error bars
in fig.~4 which were obtained using the same parameters as before.
%
\section{$pp$ collisions}

With the advent of RHIC, spin transfer reactions can be 
studied for the first time also in $pp$ scattering at c.m.s.\ energies of 
up to $\sqrt{s}=500\,\mathrm{GeV}$. In the following we will demonstrate that
such measurements would provide a particularly clean way of 
discriminating between the various conceivable sets of 
spin-dependent $\Lambda$ fragmentation functions presented above. 
For this purpose, only {\em one} polarized
beam at RHIC would be needed. It should be noted here that similar 
(and almost equally useful) measurements could be performed also in a 
possibly forthcoming experiment at DESY, HERA-$\vec{N}$.
The process we are interested in is $p\vec{p}\rightarrow \vec{\Lambda} X$
(the arrows denoting a longitudinally polarized particle) at large transverse
momentum $p_T$ of the $\Lambda$, where perturbative QCD can be safely
applied. For the time being, the required partonic helicity transfer cross 
sections, i.e., $q\vec{q}\rightarrow q \vec{q}$, $\ldots$, 
$g\vec{g}\rightarrow g \vec{g}$, are calculated only to leading order 
 accuracy. Hence we have to restrict our analysis to LO, implying the use of
LO-evolved $\Lambda$ fragmentation functions. 

The relevant differential polarized cross section can be schematically 
written as (the subscripts ``$+$'',``$-$'' below denote 
helicities)
\begin{eqnarray}
\label{eq:cross}
\frac{d\Delta \sigma^{p\vec{p}\rightarrow \vec{\Lambda} X}}{d \eta}
&\equiv& \frac{d\sigma^{pp_+\rightarrow \Lambda_+ X}}{d \eta} -
\frac{d\sigma^{pp_-\rightarrow \Lambda_+ X}}{d \eta} \\
&&\hspace*{-2.5cm} = \int_{p_T^{min}} \hspace{-0.2cm} dp_T 
\sum_{ff'\rightarrow i X} 
\int dx_1 dx_2 dz\, f^p(x_1) \times \Delta f'^p(x_2) \times
\Delta D_i^{\Lambda}(z) \times 
\frac{d\Delta \sigma^{f\vec{f'}\rightarrow \vec{i}X}}{d\eta} \; ,
\nonumber
\end{eqnarray}
the sum running over all possible LO subprocesses, and where we have 
integrated over $p_T$, with $p_T^{min}$ denoting some suitable lower 
cut-off. 

The directly observable quantity will be not the cross section in 
(\ref{eq:cross}) itself but the corresponding spin asymmetry,
defined as usual by
\begin{equation}
\label{eq:asym}
A^{\Lambda} \equiv \frac{d \Delta \sigma^{p\vec{p}\rightarrow \vec{\Lambda}X}/
d \eta}{d\sigma^{pp\rightarrow \Lambda X}/d \eta}
\end{equation}
where the unpolarized cross section 
$d\sigma^{pp\rightarrow \Lambda X}/d \eta$ is given by an 
expression similar to the one in (\ref{eq:cross}), with all $\Delta$'s
removed.

For the unpolarized parton distributions of the proton, $f^p$, appearing in 
(\ref{eq:cross}), (\ref{eq:asym}) we use the LO set of Ref.\ \cite{grv} 
throughout our calculations. 
Unless otherwise stated we use for the corresponding polarized densities
$\Delta f^p$ the LO GRSV ``standard'' scenario \cite{grsv}.
%
\begin{figure}[th]
\begin{center}
\vspace*{-1.5cm}
\epsfig{file=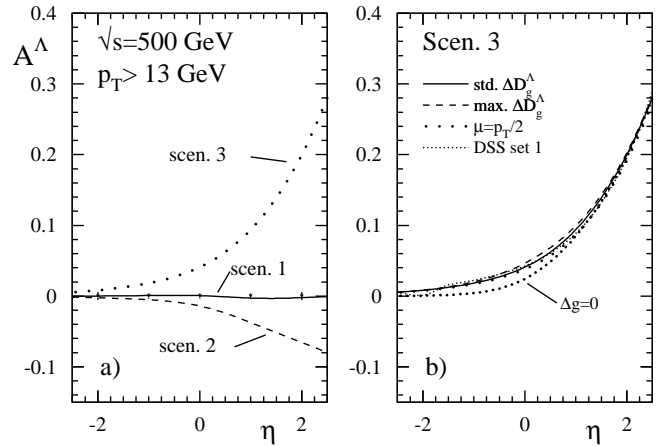,width=8.8cm}
\vspace*{-1.2cm}
\caption{ {\bf (a)} The asymmetry $A^{\Lambda}$  as a function of rapidity of the $\Lambda$ at RHIC energies
for the various sets of spin-dependent fragmentation functions.
{\bf (b)} same as for scenario 3 in {\bf (a)}, but using the ``maximal'' 
$\Delta D_g^{\Lambda}$, a hard scale $Q=p_T/2$, 
$\Delta g=0$, or the DSS1
spin-dependent parton distributions}
\vspace*{-.1cm}
\end{center}
\end{figure} 

Fig.~5(a) shows our predictions for the spin asymmetry $A^{\Lambda}$ as a
function of rapidity, calculated according to Eqs.\ (\ref{eq:asym}) and
(\ref{eq:cross}) for $\sqrt{s}=500\,\mathrm{GeV}$ and $p_T^{min}=13\,
\mathrm{GeV}$. Note that we have counted positive rapidity in the 
forward region of the {\em polarized} proton. We have used the three 
different scenarios for the $\Delta D_i^{\Lambda}$ discussed above, 
employing the hard scale $Q=p_T$. The ``error bars'' 
should give an impression of the achievable statistical accuracy for 
such a measurement at RHIC. They have been estimated by assuming a polarization $P$ of the proton beam of about 70$\%$,
a branching ratio $b_{\Lambda}\equiv BR(\Lambda\rightarrow p \pi)\simeq 0.64$,
a conservative value for the $\Lambda$ detection efficiency of 
$\epsilon_{\Lambda}=0.1$, and an integrated luminosity of 
${\cal{L}}=800\,\mathrm{pb}^{-1}$. 

The behavior of $A^{\Lambda}$ in Fig.\ 5(a) for the different
sets of polarized $\Lambda$ fragmentation functions can be easily understood 
from the fact that the process, in this particular kinematical region, is 
dominated by contributions from $u$ and $d$ quarks, so that the 
differences between the predictions in Fig.\ 5(a) are driven by the 
differences in the corresponding $\Delta D_u^{\Lambda}$ and
$\Delta D_d^{\Lambda}$. This immediately implies that the asymmetry 
has to be close to zero for scenario 1, negative for scenario 2 and 
positive and larger for scenario 3. The $\eta$-dependence is also 
readily understood: at negative $\eta$, the parton densities of the 
polarized proton are probed at small values of $x_2$ (i.e., in the 
``sea region''), where the ratio $\Delta q(x_2)/q (x_2)$ is also 
small. On the contrary, at large positive $\eta$, typical values of 
$x_2$ correspond to the valence region where the quarks are polarized much 
more strongly, resulting in an asymmetry that increases with 
$\eta$.

The results in Fig.\ 5(a) clearly demonstrate the usefulness of the
proposed 
kind of measurements to determine the polarized $\Lambda$ fragmentation 
functions more precisely. The expected statistical errors are much smaller 
than the differences in $A^{\Lambda}$ induced by the various models. 
Thus an analysis of $A^{\Lambda}$ would provide an excellent way of 
ruling out some of the presently allowed sets of spin-dependent $\Lambda$ 
fragmentation functions, {\em provided} the observed differences in 
$A^{\Lambda}$ are not obscured or washed out by the theoretical 
uncertainties inherent in this calculation. 
There are three major sources of uncertainties: the dependence of $A^{\Lambda}$
on variations of the hard scale $Q$ (implicit in Eq. (\ref{eq:cross})), which is of
particular importance since we are limited to a LO calculation, our present 
inaccurate knowledge of the precise $x$-shape and the flavor
decomposition of the polarized densities $\Delta f^p$, 
especially of $\Delta g$, and our ignorance of 
$\Delta D_g^{\Lambda}$. Fig.~5(b) gives an example of 
the scale dependence of $A^{\Lambda}$ by changing the scale from $Q=p_T$
to $Q=p_T/2$ for scenario 3. Even though $d\Delta\sigma/d\eta$ 
and $d\sigma/d\eta$ individually change by as much as a factor 2 at certain
values of $\eta$, the uncertainty almost 
cancels in the ratio $A^{\Lambda}$.  We also show in the same figure the changes in the predictions 
resulting from varying the polarized parton distributions, using the recent 
LO set 1 of Ref.\ \cite{dss}, denoted by DSS, instead of the GRSV 
\cite{grsv} one. As can be observed, the asymmetry remains practically
unchanged, and differences can only be found at the end of 
phase space (at large values of $\eta$) where the cross section becomes 
small anyway. Also, as an extreme way of estimating the impact of the  
polarized gluon distribution, we have artificially set it to 
zero $(\Delta g (x,\mu^2) \equiv 0)$.  We find that changes in our predictions 
only occur in the region of negative $\eta$, but are small in the interesting 
region $\eta>0$ where the asymmetries are larger.

Finally, in order to examine the role played by $\Delta D_g^{\Lambda}$ in our 
analysis, we have used two different approaches: the standard one for
our polarized fragmentation functions, where the polarized gluon fragmentation
function is assumed to be vanishing at the initial scale \cite{lamdis} 
and is then built up by evolution (``std. $\Delta D_g^{\Lambda}$''), and a set 
corresponding to assuming $\Delta D_g^{\Lambda} \equiv D_g^{\Lambda}$ at 
the same initial scale $\mu$ (``max. $\Delta D_g^{\Lambda}$'') 
while keeping the input quark fragmentation functions unchanged. 
As can be observed, the resulting differences are also negligible, again 
due to the fact that $u$ and $d$ fragmentation dominate.

\section{$pp$ collisions: transverse polarization}

In the case of {\em transverse} polarization  no experimental information is available. For instance, 
 the transversity densities, denoted by $\Delta_T q(x,Q^2)$
(or $h_1^q(x,Q^2)$), which are equally fundamental at leading twist 
 as the $\Delta_L q(x,Q^2)$, are 
completely unknown for the time being. 
The chiral-odd $\Delta_T q(x,Q^2)$ measures the difference of the probabilities 
to find a quark with its spin parallel to that of a transversely polarized
nucleon and of finding it oppositely polarized. 
Unlike the case of unpolarized and longitudinally polarized densities,
there is no gluon transversity distribution at leading twist, and the $\Delta_T q(x,Q^2)$
are not accessible in inclusive DIS measurements because of their
chirality properties.
In a similar way, one can define transversity fragmentation functions, 
denoted by $\Delta_T D_q^h (x,Q^2)$, to describe the fragmentation of a 
transversely polarized quark into a transversely polarized hadron. 
In view of the promising results shown before 
concerning a possible measurement of the $\Delta_L D_f^{\Lambda}(x,Q^2)$ 
in $\vec{p} p\to \vec{\Lambda} X$, 
it seems worthwhile to study this reaction for the situation of 
transverse polarization at RHIC, i.e., for 
$p^{\uparrow} p\to \Lambda^{\uparrow} X$.  
In order to be able to make sensible 
predictions for the possible spin-transfer asymmetries for this process, 
we will exploit the positivity constraints derived in \cite{js} to
constrain the involved quantities $\Delta_T q(x,Q^2)$ and 
$\Delta_T D_q^h (x,Q^2)$ in a non-trivial way.
Let us first recall that a positivity constraint at the naive 
parton model level was obtained for the $\Delta_T q(x)$, 
which reads \cite{js}
\begin{equation}
\label{eq:pos}
2|\Delta_T q(x)| \le q(x) + \Delta_L q(x) \; .
\end{equation}
An analogous positivity bound for the 
fragmentation functions of a quark $q$ into a hadron $h$ holds, namely
\begin{equation}
\label{eq:posfrag}
2|\Delta_T D_q^h (x)| \le  D_q^h (x) + \Delta_L D_q^h (x) \; .
\end{equation}
This new result is 
maintained by the QCD $Q^2$ evolution at leading order (LO).
We will use these non-trivial bounds (\ref{eq:pos}) and (\ref{eq:posfrag}) to 
constrain the unmeasured transversity parton densities $\Delta_T q(x,Q^2)$
and fragmentation functions $\Delta_T D_q^h (x,Q^2)$ in our studies
of the spin-transfer asymmetry for transversely polarized $\Lambda$ 
baryon production at RHIC below.  

The spin-transfer cross sections for the subprocesses $ff'\rightarrow 
iX'$ have been known for quite some time. 
 The cross sections for the 
transversity case were presented in \cite{handedness} 
in a form that 
also allows us to distinguish between the situations ``$S$'' and ``$N$'', i.e., when the final-state particle ``$i$'', and hence the $\Lambda$, is
transversely polarized in ($S$), or normal ($N$) to, the scattering plane.

Before we can estimate the spin-transfer asymmetry (which is usually called $D_{NN}^\Lambda$) 
for transversely polarized $\Lambda$ baryon production,
we have to specify the various different parton distributions and 
fragmentation functions involved in this calculation. 
We will use the approach of {\em saturating} the positivity 
inequalities given in Eqs.~(\ref{eq:pos}) and (\ref{eq:posfrag})
at the input resolution scale $\mu$
to constrain the unknown transversity parton densities $\Delta_T q(x,Q^2)$
and the $\Lambda$ fragmentation functions $\Delta_T D_q^\Lambda (x,Q^2)$, 
respectively. 
Such a framework is sufficient to derive a more or less rigorous 
estimate for an {\em upper bound} for the expected 
spin-transfer asymmetry $D_{NN}^\Lambda$. 
Since all relevant helicity transfer subprocess cross sections are 
available only at the Born level, we restrict ourselves also to LO for the
$Q^2$ evolutions of the involved parton density and fragmentation functions.
More precisely, for the  $\Delta_T q(x,Q^2)$ we use the unpolarized GRV \cite{grv} 
and the longitudinally polarized GRSV \cite{grsv} LO parton densities
$q$ and $\Delta_L q$, respectively, on the r.h.s.\ of Eq.~(\ref{eq:pos}).
The unpolarized and the longitudinally polarized fragmentation functions
$D_q^\Lambda$ and $\Delta_L D_q^\Lambda$ 
 serve to constrain
the transversity fragmentation functions $\Delta_T D_q^\Lambda (x,Q^2)$ via
the bound in (\ref{eq:posfrag}).

Fig.~6 shows our predictions for the spin-transfer  asymmetry 
$D_{NN}^\Lambda$ as a function of rapidity  for 
$\sqrt{s}=500\,\mathrm{GeV}$ and $p_T^{min}=13\,\mathrm{GeV}$.  
We have used the three different scenarios for the $\Delta_T D_q^{\Lambda}$ 
discussed above, employing the hard scale $Q=p_T$. 
The possibility to have negative and positive asymmetries of the same size
for each scenario reflects the freedom in the choice of the sign
for the $\Delta_T D_q^\Lambda$ and the $\Delta_T q$ in Eqs.~(\ref{eq:posfrag})
and (\ref{eq:pos}), respectively.
The ``error bars'' in Fig.~6 should give again an impression of the achievable 
statistical accuracy for such a measurement at RHIC. They have been estimated by using the same parameters as in the longitudinal case.
In Fig.~6 we have also studied the impact of one of the major theoretical 
uncertainties in a LO calculation of $D_{NN}^\Lambda$,
the dependence on variations of the a priori unknown hard scale $Q$ 
in (\ref{eq:cross}).
Luckily, it turns out that $D_{NN}^\Lambda$ depends only very weakly on
the value of the hard scale in the range $Q=p_T/2$ to $Q=2\, p_T$, as
is demonstrated for scenario 3 in Fig.~6 (very similar results hold
for the other two scenarios). 
%
\begin{figure}[th]
\begin{center}
\vspace*{-1.4cm}
\epsfig{file=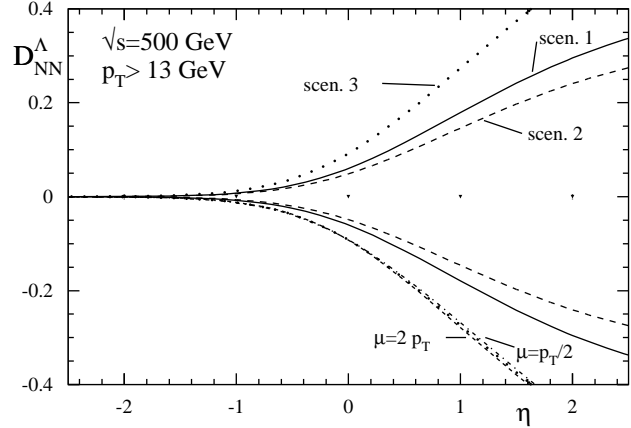,width=8.5cm}
\vspace*{-1.1cm}
\caption{  Upper bounds for the spin-transfer asymmetry $D_{NN}^{\Lambda}$   
as  functions of the rapidity of the 
produced $\Lambda$ at RHIC energies.
For ``scenario 3'' we also illustrate the typical theoretical 
uncertainty induced by a variation of the hard scale $Q$ in (\ref{eq:cross}) 
in the range $p_T/2$ to $2 p_T$.}
\vspace*{-0.2cm}
\end{center}
\end{figure} 

The results shown in Fig.~6 clearly demonstrate the usefulness of studying
also the production of transversely polarized $\Lambda$ hyperons at 
RHIC. Of course, one should keep in mind that the asymmetries presented in
Fig.~2 represent only a rough {\em upper bound} of what can be expected in an
actual measurement.
Hence the measured asymmetry  will  possibly be  considerably 
smaller with respect to our prediction, but even when reduced by a 
factor of 2 or 4, 
a measurement of $D_{NN}^\Lambda$ would still remain feasible since the 
expected statistical errors are very small.

\section{Conclusions}

Summarizing, we have shown that polarized fragmentation functions for $\Lambda$ baryons are not well determined by the available LEP data. In order to quantify the uncertainties on them we have introduced 3 different scenarios, which are particularly useful to study the sensitivity of difference observables on them.
We have first analyzed the case of Semi-Inclusive DIS, showing that measurements with unpolarized protons would allow to do a flavor decomposition of the fragmentation functions.  Furthermore, it has been shown that, counting with a polarized proton target or beam, the extraction of $\Delta s$ in the proton is not feasible from $\Lambda$ measurements, contrary to what has been claimed.
In the case of proton-proton collisions, we have shown that there are very good prospects  to measure both longitudinally and transversely polarized $\Lambda$ fragmentation functions at RHIC.

\section*{Acknowledgments}

It is a pleasure to thank J. Soffer, M. Stratmann and W. Vogelsang for enjoyable collaborations, the organizers of the EPIC99 workshop for their hospitality and M. Gutierrez for her support.

\end{document}